\documentclass[sn-mathphys,Numbered]{sn-jnl}

\usepackage{graphicx}%
\usepackage{multirow}%
\usepackage{amsmath,amssymb,amsfonts}%
\usepackage{amsthm}%
\usepackage{mathrsfs}%
\usepackage[title]{appendix}%
\usepackage{xcolor}%
\usepackage{textcomp}%
\usepackage{manyfoot}%
\usepackage{booktabs}%
\usepackage{algorithm}%
\usepackage{algorithmicx}%
\usepackage{algpseudocode}%
\usepackage{listings}%
\DeclareGraphicsExtensions{.tiff}

\newcommand{\modify}[1]{\textcolor{black}{#1}}

\begin{document}

\title[Article Title]{Development of MKIDs in the Optical and Near-infrared Bands for SPIAKID}


\author*[1,2]{\fnm{Hu} \sur{Jie}}\email{jie.hu@obspm.fr}

\author[1]{\fnm{Nicaise} \sur{Paul}}
\email{paul.nicaise@obspm.fr}

\author[1]{\fnm{Boussaha} \sur{Faouzi}}\email{faouzi.boussaha@obspm.fr}

\author[1]{\fnm{Martin} \sur{Jean-Marc}}\email{jeanmarc@caltech.edu}

\author[1]{\fnm{Chaumont} \sur{Christine}}\email{christine.chaumont@obspm.fr}

\author[1]{\fnm{Marret} \sur{Alexine}}\email{alexine.marret@obspm.fr}

\author[1]{\fnm{Reix} \sur{Florent}}\email{Florent.reix@obspm.fr}

\author[1]{\fnm{Firminy} \sur{Josiane}}\email{josiane.firminy@obspm.fr}

\author[1]{\fnm{Vacelet} \sur{Thibaut}}\email{thibaut.vacelet@obspm.fr}

\author[2]{\fnm{Pham} \sur{Viet Dung}}\email{phamvd@apc.in2p3.fr}

\author[2]{\fnm{Piat} \sur{Michel}}\email{piat@apc.in2p3.fr}

\author[1]{\fnm{Caffau} \sur{Elisabetta}}\email{Elisabetta.caffau@obspm.fr}

\author[1]{\fnm{Bonifacio} \sur{Piercarlo}}\email{piercarlo.bonifacio@obspm.fr}

\affil*[1]{\orgdiv{GEPI}, \orgname{Observatoire de Paris,PSL Université, CNRS}, \orgaddress{\street{77 avenue Denfert-Rochereau}, \city{Paris}, \postcode{75014}, \country{France}}}

\affil[2]{\orgdiv{Astroparticule et Cosmologie}, \orgname{Université de Paris, CNRS}, \orgaddress{\street{10, rue Alice Domon et Léonie Duquet}, \city{Paris}, \postcode{75013},  \country{France}}}


\abstract{SpectroPhotometric Imaging in Astronomy with Kinetic Inductance Detectors (SPIAKID) aims at designing, building, and deploying on the sky a spectrophotometric imager based on microwave kinetic inductance detectors (MKIDs) in the optical and near-infrared bands. MKIDs show a fast response and the ability to resolve photon energy compared to the conventional Charge-coupled Devices (CCDs). In this paper, we present the design and simulation of the MKID arrays for SPIAKID. The detectors consist of four arrays with each array of 20,000 lumped-element pixels, and each array will be read with 10 readout lines. 
The meander material of the resonators is trilayer TiN/Ti/TiN to have better uniformity of the critical temperature across the array. \modify{We also present the measurement result for a test array with $30\times30$ pixels  
which is a subset of the designed 2000-pixel array to verify the design and fabrication}. 
\modify{
The current measured best energy resolving power $R = E/\Delta E$ is 2.4 at $\lambda = 405~\text{nm}$ and the current medium R is around 1.7. We have also observed the response of the TiN/Ti/TiN is much smaller than expected.}
}




\maketitle

\section{Introduction}\label{Introduction}

SPIAKID is an ERC-funded project that aims to open the way to a new class of wide-range, wide-field, high-efficiency, and high-angular-resolution MKIDs-based spectro-imagers via a demonstrator instrument at NTT telescope in Chile. The primary objective of the project is to perform a detailed study of the stellar populations of at least one ultra-faint dwarf galaxy\cite{Willman2005, Willman2005_1, Bouwens2012, Bechtol2015} (UFD) in the local group. Other science cases are the follow-up observations of sources of gravitational waves\cite{Ligo2016} and afterglows of gamma-ray bursts (GRBs)\cite{Piran2005}, characterization of the minor bodies of the solar system, and detection of exoplanet transits and exoplanet transit spectroscopy\cite{Mandell2013}.

MKIDs show a great advantage\cite{Mazin2013, Meeker2018} over CCD cameras for their intrinsic energy resolution as well as the ability to record the arrival time of the photons. 
\modify{This means that for any object in the Field-of-View, one records the number of photons
arrived over a given time interval and for any chosen wavelength bin, that is a spectrum, without
the need to disperse the light with a prism or grating. The resolving power of such
spectrum, $E/\Delta E$, is dictated by the performance of the detector.}
SPIAKID aims at a Field-of-View of 2' x 2' in the sky. This can be achieved with four MKID arrays of 20,000 pixels each on the focal plane.
Due to budgetary restrictions, we shall equip the focal plane with four arrays, but we shall read only two arrays with 10 feed lines each.
Each feedline will read out 2000 pixels. The wavelength range covered by our detectors will cover the optical and near-infrared (0.4 $\rm\mu m$ to 1.6 $\rm\mu m$).


An MKID pixel in the optical band and the near-infrared band is a superconducting resonator usually consisting of an interdigital capacitor and a meander line. Each pixel has a unique resonance frequency. All the pixels share the same meander design.  Multiplexing is realized by tuning the capacitance by changing the finger length of the prototype interdigital capacitor. The resonance frequency spacing is usually on the order of 2~MHz. 

The meander is usually made of superconductors with higher normal resistivity, such as TiN\cite{Gao2012, Nicaise2022_1, Boussaha2023}, TiN/Ti/TiN\cite{Vissers2013}, PtSi\cite{Szypryt2017}, Hafnium\cite{Zobrist2019}, Hafnium/Indium\cite{Zobrist2022} and $\beta$-Tantalum\cite{Kouwenhoven2023}, which is quite challenging to keep high resistivity and high film quality. TiN/Ti/TiN for better uniformity over the wafer, relative ease of fabrication, as well as high quality. 

Designing a 2000-pixel array is also not easy. First, as the capacitance becomes smaller, the length change in the finger of the interdigital capacitor becomes smaller, eventually becoming less than $1~\rm\mu \text{m}$, which is difficult for fabrication, especially with regular uv lithography. The second is frequency collision originating from the fabrication uncertainty, especially from the size of the meander. 

In this paper, we introduce the design of the MKID array for SPIAKIDs with an increasing gap in the capacitor to reduce the resonance frequency sensitivity to finger length as the resonance frequency increases. 

\modify{
We will also present the measurement result of an MKIDs array with $30\times30$ pixels, which is a subset of the designed 2000 pixels array. This resonance frequency ranges from 4-8~GHz, with a frequency spacing of about 4~MHz, which is chosen based on the limited internal quality factor of our current fabrication. The array is a key step in verifying the design and the fabrication procedure for the full array for SPIAKID. 
}

\section{MKIDs Design and Simulation}\label{sec: MKIDs design and simulation}

\begin{figure}[ht]
    \centering
    \includegraphics[width =\textwidth]{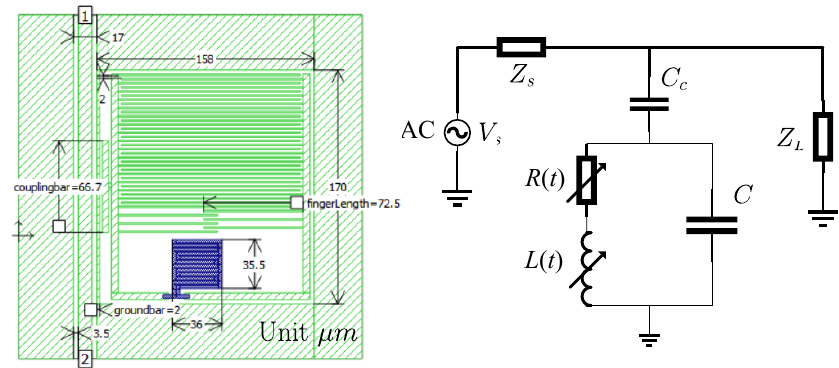}
    \caption{One of the designed MKID pixels in the array with resonance around 6~GHz and the equivalent circuit of MKIDs}
    \label{fig:pixel design}
\end{figure}

The design goal of the MKID array for SPIAKID is to design an MKID array with resonance frequencies ranging from 4 to 8~GHz with a length variation in the finger of each capacitor greater than $1~\rm\mu m$ and a resonance spacing of about 2~MHz, making it possible to fabricate the MKID array in ordinary lithography. Here, we gradually increase the gap between the fingers of the capacitor to reduce the capacitance sensitivity to the length of the finger. One of the designed MKIDs is shown in Fig.\ref{fig:pixel design}.  

We select the material of the meander to be TiN/Ti/TiN\cite{Vissers2013} mainly to improve the uniformity of the film across the wafer. 
The film of the meander is made of TiN/Ti/TiN with a critical temperature of 1.75~K and thickness to be 10/10/10 deposited on a sapphire substrate. The resistivity of the trilayer film is about $93~\mu \Omega \cdot \text{cm}$, corresponding to kinetic inductance\cite{Gao2008} $L_k \approx 24.5~pH/\square$. The meander is a double-folded meander to reduce the crosstalk between the pixels\cite{Noroozian2012}. The size of the meander is $36\times 36~\rm\mu m^2$, to accommodate the optics from the telescope, which includes a microlens set that is about 0.7~mm above the MKIDs array. The width of the meander line is $2.5~\mu$m and the gap between the meander line is $0.5~\mu $m. The distance between the adjacent MKID pixels is $180~\rm\mu m$, which corresponds to $0.45^{"}$ on the sky based on our current optical design. 

The width of the fingers in the capacitor is fixed at $1~\rm\mu m$, while the spacing between them changes from $1~\rm\mu m$ to $4.5~\rm\mu m$ when the resonance frequency changes from 4 to 8~GHz.

We couple the resonator to the feedline with a coupling bar. \modify{The coupling quality factor of the MKIDs scales with the coupling capacitor as follows}\cite{Noroozian2012_1} 
\begin{align}\label{eqn: Qc}
    Q_c = \frac{2}{Z_0L_0C_c^2\omega_r^3} 
\end{align}
where $Z_0$ is the characteristic impedance of the feedline, $L_0$ is the total inductance in the resonator, $C_c$ is the coupling capacitor to the feedline and $\omega_r = 2\pi f_r$ is the angular frequency with $f_r$ the resonance frequency. 
And $C_c$ can be further expressed as a parallel connection of the coupling capacitor $C_{cc}$ and the parasitic capacitor to the ground $C_{c0}$ as 
\begin{align}\label{eqn: Cc}
    C_c = C_{cc} + C_{c0}
\end{align}
$C_{cc}$ can be adjusted by tuning the length of the coupling bar by simulating the resonator with different lengths of the coupling bar and fitting the $S_{21}$ of the resonator with 
\begin{align}\label{eqn: s21}
    S_{21} =  1 - \frac{Q_L/Q_c}{1+2jQ_L\delta x}
\end{align}
where $Q_L$ is the quality factor of the resonator and $\delta_x = (f-f_r)/f_r$ is the fraction of frequency shift to the resonance frequency $f_r$.

\begin{figure}
    \centering
    \includegraphics[width = \textwidth]{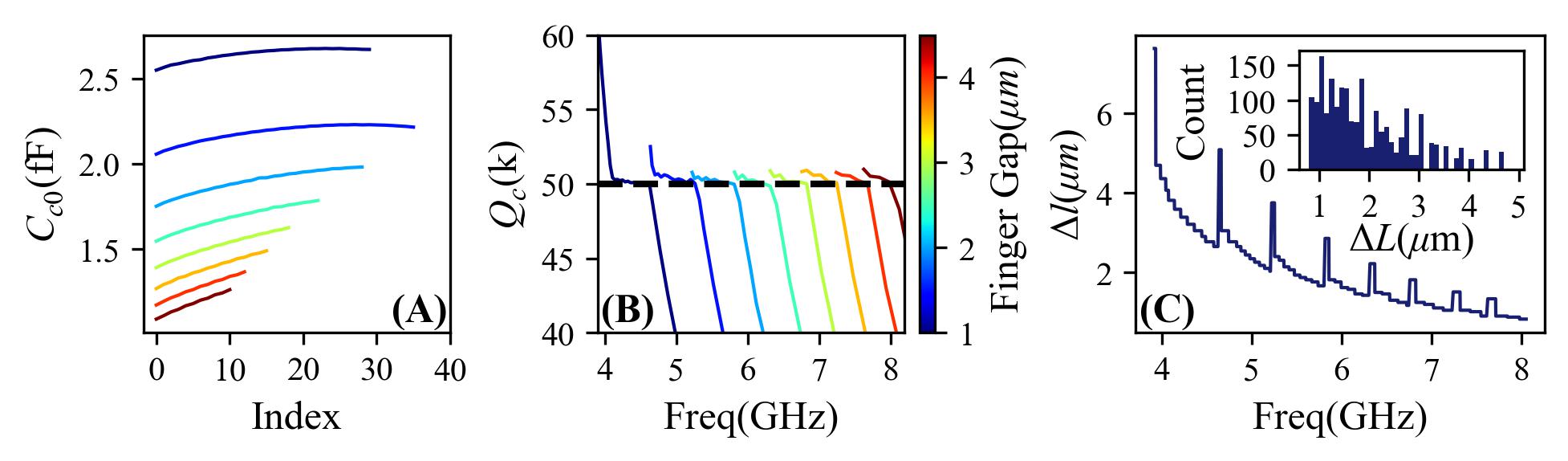}
    \caption{(A): Simulated $C_{c0}$ versus the number of the fingers in the capacitor that has been shorted. (B): Tuned $Q_c$ versus the resonance frequency. (C): The change of finger length $\Delta l$ versus the resonance frequency. The inset shows the statistics of $\Delta l$ for the 2000-pixel array.}
    \label{fig: Tunning QC}
\end{figure}

$C_{c0}$ corresponds to the coupling capacitance when $C_{cc} = 0$ and tends to change with resonance frequency. To tune the $Q_c$ to be around the desired value, which is 50,000 in our case, $C_{c0}$ is simulated with different capacitance tuned by the finger length in the interdigital capacitor as is shown in Fig.~\ref{fig: Tunning QC}-(A). $C_{c0}$ tends to increase and saturate, which means that $Q_c$ cannot be tuned further. To solve this problem, we adjust the gap width in the interdigital capacitor and the ground bar width, as shown in Fig.~\ref{fig:pixel design} to tune the $C_{c0}$. We further tune $Q_c$ by tuning $C_{cc}$, which can be obtained by sweeping the coupling bar length. Once $C_{c0}$ and $C_{cc}$ are obtained, we interpolate the length of the coupling bar to obtain the desired $Q_c$ which is shown in Fig.~\ref{fig: Tunning QC}-(B). Finally, we linearly interpolate each finger to obtain the array. The change in finger length $\Delta l$ is shown in Fig.~\ref{fig: Tunning QC}-(C), and the statistics of $\Delta l$ are shown in the inset of Fig.~\ref{fig: Tunning QC}-(C), which shows that most of $\Delta l$ is larger than $1~\rm\mu m$. 

\section{MKIDs Characterization}
\begin{figure}
    \centering
    \includegraphics[width = \textwidth]{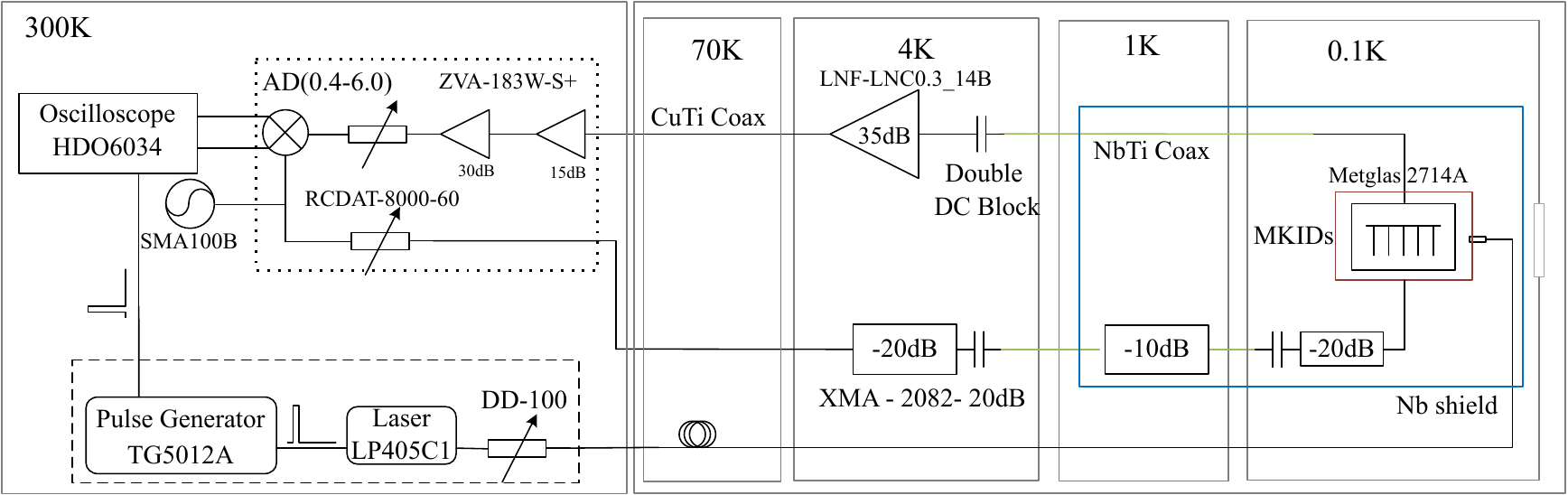}
    \caption{Detailed measurement setup for the MKIDs in the ADR}
    \label{fig: Measurement Setup}
\end{figure}

The MKIDs is measured in an adiabatic demagnetization refrigerator (ADR) at Laboratoire Astroparticule \& Cosmologie (APC). \modify{The detailed measurement setup is shown in Fig.~\ref{fig: Measurement Setup}. 
The stray magnetic field is shielded by a niobium cylinder of 1.5\,mm thickness and sheets of metglas 2714a around MKIDs. The MKIDs are readout by a standard homodyne mixing scheme. The input signal is generated by a signal generator attenuated 20 dB, 10 dB, and 20 dB on 4 K, 1 K, and 100 mK to reduce the thermal noise. The output signal from MKIDs is first amplified by an LNA on the 4K stage and amplified further by two room-temperature amplifiers, which is the main source for the readout noise The signal is down-converted to DC by an IQ mixer and then sampled by an oscilloscope. Two double DC blocks are placed between the 4K and 1K stages for the operation of the heat switch in the cryostat. The MKIDs array is illuminated by an optical fiber that is placed 35 mm above the pixels. The laser is modulated by a 250 Hz pulse from the pulse generator. The output power of the laser is estimated to be a few pW outside the cryostat, attenuated by a digital step attenuator. The pulse response of the MKID is sampled by an oscilloscope (HDO6034) at 100 MHz. The $S_{21}$ is measured by replacing the IQ mixer with a VNA.} It is the same measurement setup used in our previous publications\cite{Hu2020, Boussaha2023}. 

\begin{figure}[h]
    \centering
    \includegraphics[width = 
    \textwidth]{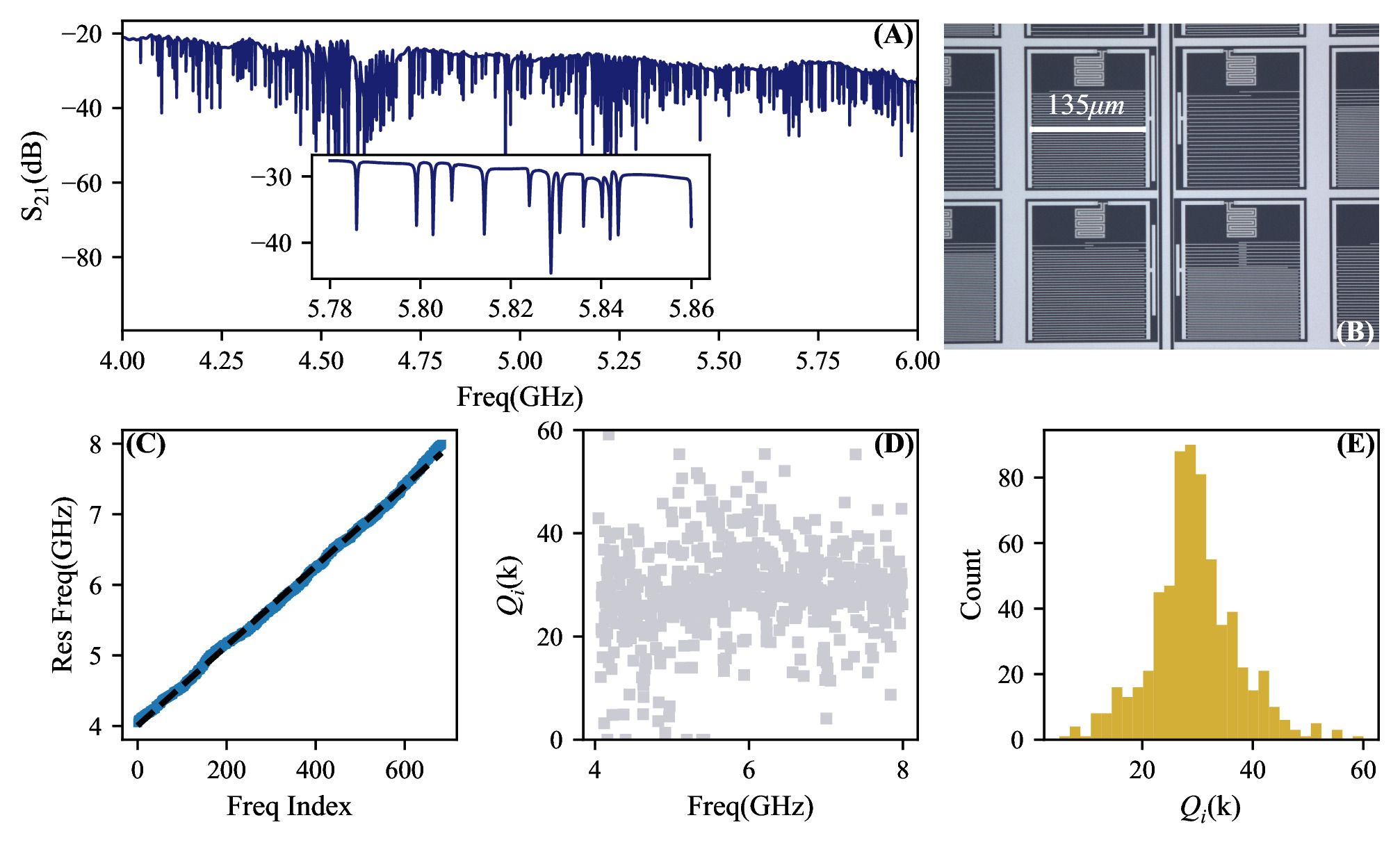}
    \caption{Measured $S_{21}$ for the MKIDs array and its statistics. (A): The measured $S_{21}$ between 4-6~GHz. The inset shows the transmission between 5.78-5.86~GHz. (B): Photon of measured MKIDs chp. (C): The resonance frequency versus the resonance index. (D): The fitted $Q_i$ versus resonance frequency. (E): The statistics of the $Q_i$}
    \label{fig:S21 and its statistics}
\end{figure}

The Measured MKIDs array has $30\times30$ pixels, which is a subset of the designed 2000-pixel array. The MKIDs were fabricated by magneton sputtering in the clean room in Paris Observatory. The picture of the fabricated MKIDs is shown in Fig.~\ref{fig:S21 and its statistics}-(B).
The measured $S_{21}$ of the array is shown in Fig.~\ref{fig:S21 and its statistics}-(A). The resonance frequency starts from 4~GHz, which is in good agreement with the simulation. The resonance frequency versus the resonance index shown shows quite a good linearity, which indicates the cross-coupling between the pixels is acceptable\cite{Noroozian2012}, as is shown in Fig.~\ref{fig:S21 and its statistics}-(C). We have fitted $Q_i$ with Eq.~(\ref{eqn: s21}) in Fig.~\ref{fig:S21 and its statistics}-(D) and it doesn't show significant frequency dependence. The yield of the array is around 75\%. The median internal quality factor $Q_i$ is around $30,000$ and it doesn't show significant frequency dependence, as is shown in Fig.~\ref{fig:S21 and its statistics}-(D). 

\begin{figure}[h]
    \centering
    \includegraphics[width = \textwidth]{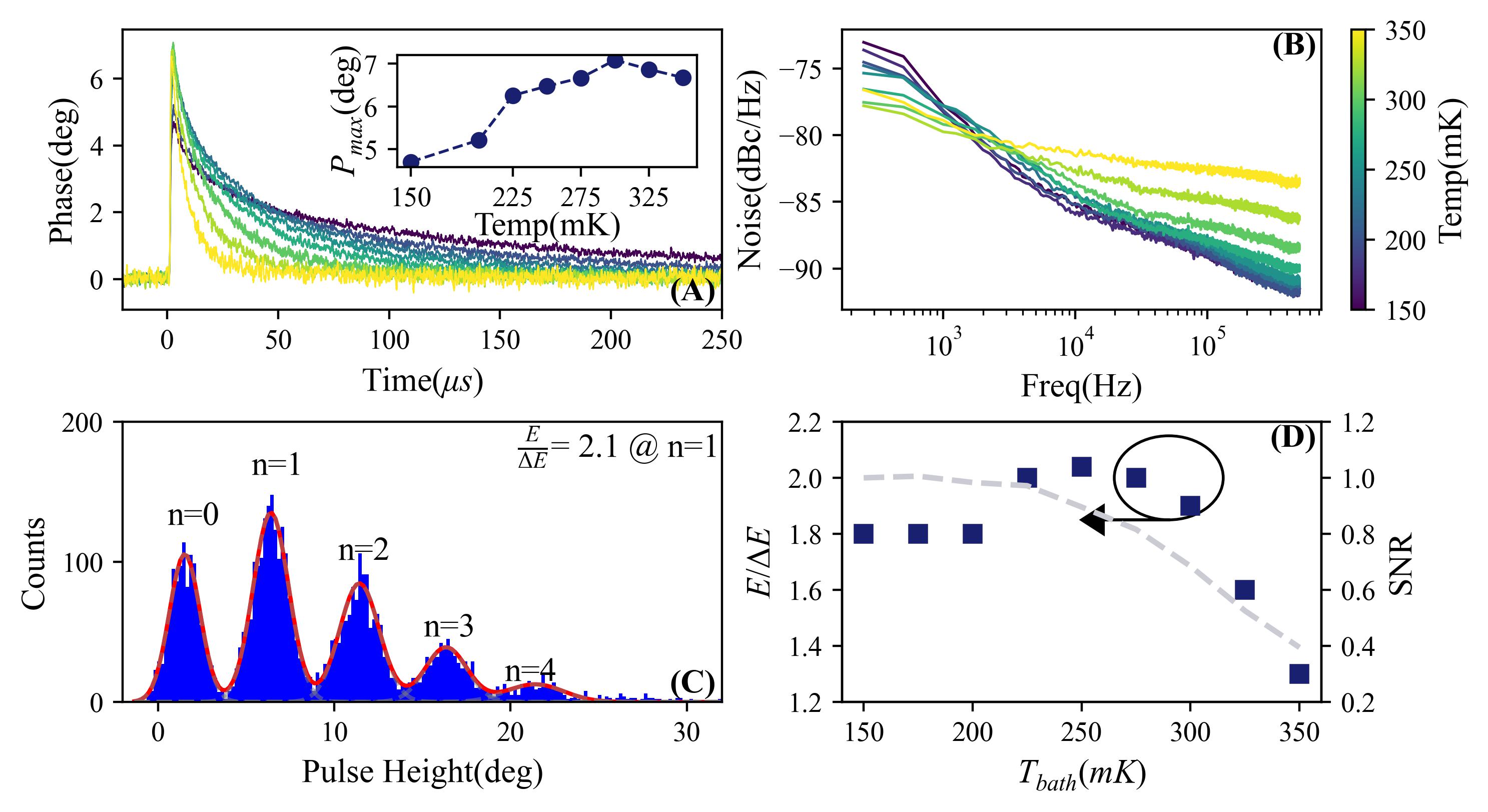}
    \caption{(A): Averaged pulse of single photon response at 405~nm at different $T_{bath}$. The inset shows the maximum of the pulse response versus $T_{bath}$. (B): Noise spectrum measured at different $T_{bath}$. (C): The pulse statistics at $T_{bath} = 250$~mK. (D): Fitted energy resolution and SNR versus $T_{bath}$.}
    \label{fig:pulse statistics and noise}
\end{figure}

We show the single photon performance of a pixel with a resonance frequency of around 5.8~GHz in Fig.~\ref{fig:pulse statistics and noise}. The $Q_i$ of the pixel is around 53,000. Fig.~\ref{fig:pulse statistics and noise}-(A) shows the single photon phase response of the pixel at different bath temperatures ($T_{bath}$) that is averaged from the full width at half maximum (FWHM) in the pulse statistics and the inset shows the pulse maximum that increases about 50\% from $T_{bath} = 150$~mK to $T_{bath} = 300$ ~mK. The noise spectrum at different temperatures is shown in Fig.~\ref{fig:pulse statistics and noise}-(B).

\modify{The pulse statistics is shown in Fig.~\ref{fig:pulse statistics and noise}-(C), which shows the $E/\Delta E$ to be around 2.1 @405~nm. The 1-$\sigma$ width of the 1-photon peak and the 0-photon peak is 1.1 deg and 0.8 deg respectively, which indicates the noise from readout system is not significant. } It can be seen from Fig.~\ref{fig:pulse statistics and noise}-(D) that energy resolving power increases a bit as $T_{bath}$ increases, which can be attributed to the increasing signal-to-noise ratio (SNR) and a reduction in the noise of the two-level system\cite{Gao2008_1, Hu2021}, as indicated in Fig.~\ref{fig:pulse statistics and noise}-(C). This phenomenon has also been observed in single-layer TiN MKIDs\cite{Boussaha2023}, \modify{but much less significant. The $E/\Delta E$ is estimated to be around 1.2-1.4 for the same pixel based on our measurement on another array with the same design. } 

\modify{
It should also be noted that the $E/\Delta E$ we obtained is comparable with those published results\cite{Guo2017}, considering the volume of our meander $V=32.4~\mu \text{m}^3$. 
}
\modify{
The response of the MKIDs is much smaller than what we expected. The expected response of the MKIDs is\cite{Martinez2019}
\begin{align}
    \phi_{qp} = \frac{\alpha S_2Q_L}{N_0 V\Delta } \cdot \frac{\eta E}{\Delta}
\end{align}
where $\alpha\approx1$ is the fraction of the kinetic inductance. $S_2 = 2.73$ is so-called the Mattis-Bardeen factor\cite{Zmuidzinas2012} with $f_0 = 5.8~\text{GHz}$, $N_0$ is the single spin density on the Fermi level, $\eta$ is the pair-breaking efficiency, $E=3.06~\text{eV}$ is the photon energy of a 405~nm photon, and $\Delta\approx 1.76k_B T_c$ is the energy gap of the superconductor. }

\modify{
If we assume $\eta = 0.6$ and $N_0 = 6.0\times 10^{10} \text{eV}^{-1}\mu m^{-3}$, which is the value for a single layer TiN\cite{Kardakova2013}, with $Q_L \approx 14000$, the estimated $\phi_{qp}\approx 30~\text{deg}$, about 5 times higher than the value we have measured. In this case, there are two possible reasons. The first is the $N_0$ for the TiN/Ti/TiN film could be much higher than the single layer TiN. The other possible reason is the pair-breaking efficiency $\eta$ is much smaller in the TiN/Ti/TiN film due to the different quasi-particle energy in different layers. 
}

\begin{figure}[ht]
    \centering
    \includegraphics[width= \textwidth]{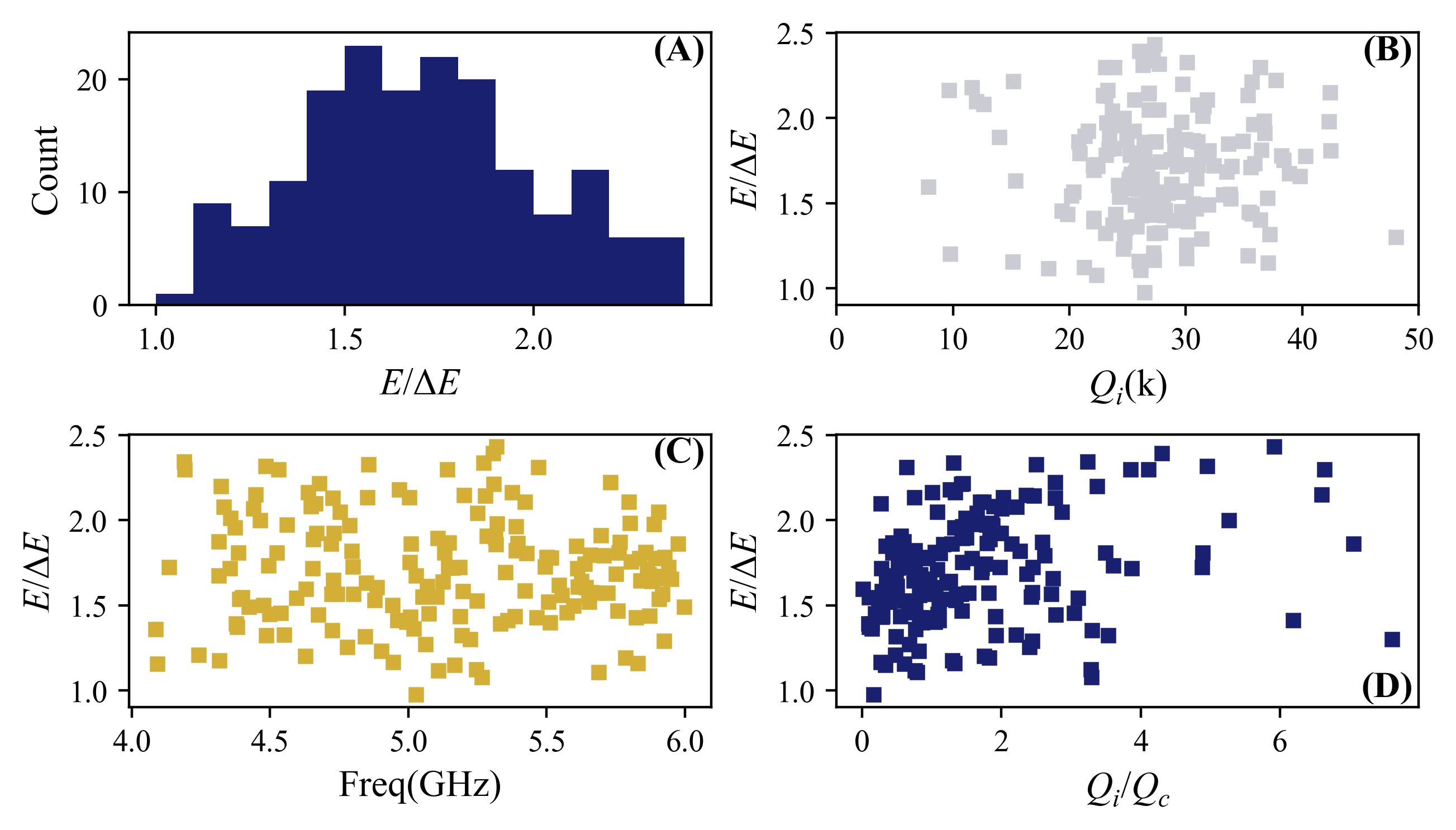}
    \caption{(A): Statistics of $E/\Delta E$ for resonators from 4.3~GHz to 6~GHz, \modify{around 330 pixels are measured.} (B-D):$E/\Delta E$ versus $Q_i$, $f_0$ and $Q_i/Q_c$ }
    \label{fig:statistics of energy resolution}
\end{figure}

We show the statistics of the energy-resolving power of the pixels with resonance frequency between 4-6~GHz \modify{measured at 250~mK}. About 330 resonators have been measured \modify{individually}, about 70\% of which show a single photon response. The median $E/\Delta E$ is about 1.7. We do not observe significant dependence between $E/\Delta E$ and the internal quality $Q_i$ as well as $f_r$. $E/\Delta E$ tends to increase when the ratio between $Q_i$ and $Q_c$ increases, which is shown in Fig.~\ref{fig:statistics of energy resolution}-(D), which is reasonable as when $Q_i>Q_c$, the readout noise in the system tends to be suppressed as the ratio of resonance circle increases. 

\modify{
The main reason for this low $E/\Delta E$ is that the response of the TiN/Ti/TiN film is unexpectedly low. We would like to reduce the size to 25x25 $\mu m^2$ or even 15x15 $\mu m^2$. The second is to optimize the film quality. Currently, the $T_c$ of the TiN/Ti/TiN is around 1.75 K. We would further optimize the film to have a $T_c=1.2-1.4$~K. In this case, we hope
we can get an improvement of $E/\Delta E$ with a factor of 3-4.}

\modify{We stress, however, that even this low $E/\Delta E$ is sufficient for the main science goal
of SPIAKID, namely the characterization of the stellar populations of
Ultra Faint Dwarf galaxies (UFDs). We reached this conclusion after analyzing
synthetic fluxes computed from model stellar atmospheres.
The main conclusion is that stellar parameters can be extracted from a SPIAKID
spectrum even if $E/\Delta E$ is as low as currently afforded by our detectors,
provided it is known for any given wavelength.
We used theoretical stellar fluxes, computed from one-dimensional model stellar atmospheres in 
hydrostatic equilibrium, computed with a resolving power $R=E/\Delta E = 200$.
We used fluxes of two models with the same effective temperature and gravity, but different concentrations of 
elements heavier than He, and 
scaled the fluxes to the absolute magnitude predicted by theoretical isochrones
of two different ages. Such a combination is what you can expect to find
among the stellar populations of UFDs. 
We verified that the monochromatic magnitude difference between the two
fluxes is about 0.2\,magnitudes, both for the original fluxes and for
those degraded to R=2.5. Hence it shall be, in principle, possible
to derive stellar parameters from the observed SPIAKID spectra, by fitting
theoretical spectra. This requires, of course, that the resolving power be known
at any given wavelength. This information shall be provided by the SPIAKID
calibration plan.
}

\section{Conclusion}
We have designed, fabricated, and characterized the TiN/Ti/TiN MKID array for SPIAKID. The current results show that the yield of the current design is around 75\% out of 900 pixels and around 70\% of measured pixels show a single photon response at 405~nm. The optimum $E/\Delta E$ is around 2.4 and the median is around 1.7. The TiN/Ti/TiN MKIDs show a similar response with bath temperature compared to the single-layer TiN MKIDs. \modify{The response of TiN/Ti/TiN is around 5 times smaller than what we expected. }

\section{Acknowledgement}
This work is supported
by the European Research Council (ERC) through Grant  835087 (SPIAKID) and UnivEarhS Labex
program.

\bibliography{sn-bibliography.bib}

\end{document}